\documentclass[apjl]{emulateapj}
\usepackage{amsmath}
\usepackage{multirow}
\usepackage{epstopdf}

\def \aap{A\&A}
\def \apjl{ApJ}

\def \apjs{ApJS}

\shorttitle{DIB 8620 in RAVE}
\shortauthors{Kos et al.}

\renewcommand{\textbf}{}

\begin{document}

\title{Diffuse interstellar band at 8620 \AA\ in RAVE: A new method for detecting the diffuse interstellar band in spectra of cool stars}

\author{J. Kos\altaffilmark{1}, T. Zwitter\altaffilmark{1,2}, E. K. Grebel\altaffilmark{3}, O. Bienayme\altaffilmark{4}, J. Binney\altaffilmark{5}, J. Bland-Hawthorn\altaffilmark{6}, K. C. Freeman\altaffilmark{7}, B. K. Gibson\altaffilmark{8}, G. Gilmore\altaffilmark{9}, G. Kordopatis\altaffilmark{9}, J. F. Navarro\altaffilmark{10}, Q. Parker\altaffilmark{11,12,13}, W. A. Reid\altaffilmark{11,12}, G. Seabroke\altaffilmark{14}, A. Siebert\altaffilmark{4}, A. Siviero\altaffilmark{14,16}, M. Steinmetz\altaffilmark{16}, F. Watson\altaffilmark{13}, R. F. G. Wyse\altaffilmark{17}}

\affil{$^1$Faculty of Mathematics and Physics, University of Ljubljana, Jadranska 19, 1000 Ljubljana, Slovenia; \\ e-mail: janez.kos@fmf.uni-lj.si}
\affil{$^2$Center of Excellence SPACE-SI.A\v{s}ker\v{c}eva cesta 12, 1000 Ljubljana, Slovenia}
\affil{$^3$Astronomisches Rechen-Institut, Zentrum f\"ur Astronomie der Universit\"at Heidelberg, M\"onchhofstra\ss e 12 -- 14, D-69120 Heidelberg, Germany}
\affil{$^4$Observatorie astronomique de Strasbourg, Universit\'e de Strasbourg, 11 rue de l'Universit\'e, F-67000 Strasbourg, France}
\affil{$^5$Rudolf Peierls Centre for Theoretical Physics, Keble Road, Oxford =X1 3NP, UK}
\affil{$^6$Sydney Institute for Astronomy, School of Physics A28, University of Sydney, NSW 2008, Australia}
\affil{$^7$Research School of Astronomy \& Astrophysics, Australian National University, Canberra, Australia}
\affil{$^8$Jeremiah Horrocks Institute, University of Central Lancashire, Preston, PR1 2HE, United Kingdom}
\affil{$^9$Institute of Astronomy, University of Cambridge, Madingley Road, Cambridge, CB3 0HA, UK}
\affil{$^{10}$Senior ClfAR Fellow. University of Victoria, Victoria BC, Canada V8P 5C2}
\affil{$^{11}$Department of Physics and Astronomy, Macquarie University, Sydney, NSW 2109, Australia}
\affil{$^{12}$Centre for Astronomy, Astrophysics and Astrophotonics, Macquarie University, Sydney, NSW 2109, Australia}
\affil{$^{13}$Australian Astronomical Observatory, PO Box 915, North Ryde, NSW 1670}
\affil{$^{14}$Mullard Space Science Laboratory, University College London, Holmbury St Mary, Dorking, RH5 6NT, UK}
\affil{$^{15}$Department of Physics and Astronomy, Padova University, Vicolo dell'Osservatorio 2, I-35122 Padova, Italy}
\affil{$^{16}$Leibniz-Institut f\"ur Astrophysik Potsdam (AIP), An der Sternwarte 16, 14482 Potsdam, Germany}
\affil{$^{17}$Johns Hopkins University, Homewood Campus, 3400 N Charles Street, Baltimore, MD 21218, USA}

\begin{abstract}
Diffuse interstellar bands are usually observed in spectra of hot stars, where interstellar lines are rarely blended with stellar ones. The need for hot stars is a strong limitation in the number of sightlines we can observe and the distribution of sightlines in the Galaxy, as hot stars are rare and concentrated in the Galactic plane. We are introducing a new method, where interstellar lines can be observed in spectra of cool stars in large spectroscopic surveys. The method is completely automated and does not require prior knowledge of the stellar parameters. If known, the stellar parameters only reduce the computational time and are not involved in the extraction of the interstellar spectrum. The main step in extracting interstellar lines is a construction of the stellar spectrum, which is in our method done by finding other observed spectra that lack interstellar features and are otherwise very similar to the spectrum in question. Such spectra are then combined into a single stellar spectrum template, which matches the stellar component in an observed spectrum. We demonstrate the performance of this new method on a sample of 482,430 spectra observed in RAVE survey. \textbf{However, many spectra have to be combined (48 on average) in order to achieve a S/N ratio high enough to measure the DIB's profile, hence limiting the spatial information about the ISM.} Only one \textbf{strong} interstellar line is included in the RAVE spectral range, a diffuse interstellar band at 8620 {\AA}. We compare its equivalent width with extinction maps and with Bayesian reddening, calculated for individual stars, and provide a linear relation between the equivalent width and reddening. Separately from the introduced method, we calculate equivalent widths of the diffuse interstellar band in spectra of hot stars with known extinction and compare all three linear relations with each other and with relations from the literature.
\end{abstract}


\section{Introduction}

The origin of diffuse interstellar bands (DIBs) are one of the longest standing problems of astronomical spectroscopy \citep{herbig95, sarre06}. Discovered in 1919 \citep{heger22} as single lines in spectra of binary stars with otherwise doubled lines, they have yet unknown carriers. More than 400 DIBs are known to date \citep{hobbs09}. They are mostly found in the optical and near infra-red spectral bands, with the DIB with the longest wavelength discovered at 1.793~$\mu m$ \citep{geballe11}. DIBs were also observed in nearby galaxies \citep{vidal87, cox07, cordiner08, cordiner08b} and at cosmological distances \citep[e.g.][]{york06}, but most of the studies rely on high resolution and high S/N spectra of hot stars in our galaxy. Because the DIBs are weak (the strongest one at 4428 {\AA} having a typical equivalent width of 2~{\AA} in a {E(B-V)=1} sightline) and easily blended with stellar lines, high S/N spectra of hot stars are most appropriate for studying DIBs. Therefore most of the surveys include few thousand stars at most \textbf{\citep{snow77,loon13}}, or only around a hundred stars, if weaker DIBs are observed \textbf{\citep[e.g.][]{friedman11}}. Only in recent years the surveys capable of detecting DIBs in more stars are becoming reality\textbf{, like RAVE \citep{rave06, rave08, rave11, kordopatis13}, SDSS-III \citep{eisenstein11}, GOSSS \citep{maiz11}, Gaia-ESO \citep{gilmore12}, Gaia \citep{gaia12}, Hermes-GALAH \citep{galah12} or LAMOST \citep{deng12}.} Observations of 100,000s of stars bring new possibilities to the study of DIBs, to map the distribution of carriers in the Galaxy and to search for peculiar environments with strange DIB properties. All this can contribute to the big goal of identifying the carriers. Even without the knowledge of the carriers, DIBs can be used to trace unobserved or hard to observe properties of the ISM toward the stars in a spectroscopic survey. All \textbf{extensively studied} DIBs correlate at least vaguely with reddening and H\textsc{I} abundance \citep{herbig95}, ratios of different DIB strengths correlate with the UV radiation field \citep{krelowski92} and widths of some DIBs correlate with H$_2$ abundances \citep{gnacinski13}. For surveys focused on the study of the kinematics of stars, like RAVE,  precise distances are needed and they can only be calculated when the reddening is known. DIB at 8620~{\AA}, which seems to correlate well with reddening \citep{munari08}, could be a good alternative to photometrically measured reddening.

If DIBs are to be observed on a large scale, covering the substantial fraction of the sky, they need to be detectable in spectra of cool stars, because hot stars are not numerous enough and are almost exclusively found in the Galactic plane. The only method described in the literature uses synthetic spectra as templates for the stellar spectrum \citep{chen12}.

The Radial Velocity Experiment (RAVE)\citep{rave06, rave08, rave11, kordopatis13} is a spectroscopic survey with the goal to measure radial velocities and basic stellar atmosphere parameters for almost 500,000 stars using the Six Degree Field multi-object spectrograph on the 1.2 m UK Schmidt Telescope of the Australian Astronomical Observatory. The spectral window is centred on the near infra-red Ca~\textsc{ii} triplet (8410\AA\ - 8795\AA) and includes one strong diffuse interstellar band (DIB) at 8620 \AA. This DIB is named in the literature as DIB 8620 or DIB 8621, depending on the authors' preference. It is also known as the ''Gaia DIB'', because it is included in the spectral range of the Gaia radial velocity spectrograph. In this paper, it will be named DIB 8620. 

DIB 8620 in RAVE was previously studied in \citet{munari08}, where 68 hot stars from RAVE had this DIB measured and compared to the reddening. A very good correlation was found between the DIB's equivalent width and the reddening. \textbf{\citet{munari08} reports another DIB at 8648~{\AA}, that is much weaker than the DIB 8620. We were able to detect this DIB in only few spectra and in only few sightlines, but were unable to measure it's profile accurately enough to perform any further analysis.}

In addition, the correlation of DIB 8620 with reddening was studied in \citet{munari00} and \citet{sanner78}. \citet{cox11} observed polarisation of $\sim$30 DIBs, including 8620 and concluded that DIB 8620 is not associated with dust grains. DIB 8620 was also detected in some spectra of the SDSS survey \citep{yuan12}, but no further analysis was made. In the aspect of identifying DIB carriers, DIB 8620 is mentioned in \citet{salama99}, because a band of the tetracene ion (C$_{18}$H$_{12}^+$) lies at 8648 \AA. However the two bands are not connected, as their line profiles do not match.

In this paper we present a new method for the extraction of the DIB profile from spectra of cool stars. By ''cool'' we mean effective temperatures below 7000 K, because a simpler method can be used for stars with T$\gtrsim$7000 K, where this DIB is not blended with stellar lines. RAVE has observed so many spectra that the observed spectra themselves can be combined into a stellar spectrum template. The observed spectrum can be divided by such template to extract the DIB spectrum. We also study the correlation of the equivalent width of DIB 8620 with reddening. Section \ref{sec:data} presents the RAVE database and describes our new method of DIB extraction. Because the method has not been described previously, we present 3 tests in Section \ref{sec:tests}, where we demonstrate that the extraction process is independent of stellar parameters. Section \ref{sec:corr} presents the correlations of DIB 8620 with reddening. Section \ref{sec:conclusion} concludes with a summary of results. 

\section{Data analysis}
\label{sec:data}
\subsection{Data overview}

For all the analysis presented in this paper we used spectra of the data release 4 (hereafter DR4), which includes 482,430 spectra of 439,503 objects. Stellar parameters are calculated for 410,837 objects and photometric distances for 397,783 of these objects. DR4 \citep{kordopatis13} is based on the pipeline presented in \citet{kordopatis11},  and is more than 5 times bigger than data release 3. It includes  updated parameters compared to the older data releases, which are now more accurate, since the spectral degeneracies and the 2MASS photometry are better taken into account. These improved parameters lead naturally to improved accuracy on the stellar distances. 

\begin{figure*}[!t]
\centering
\includegraphics[width=\columnwidth]{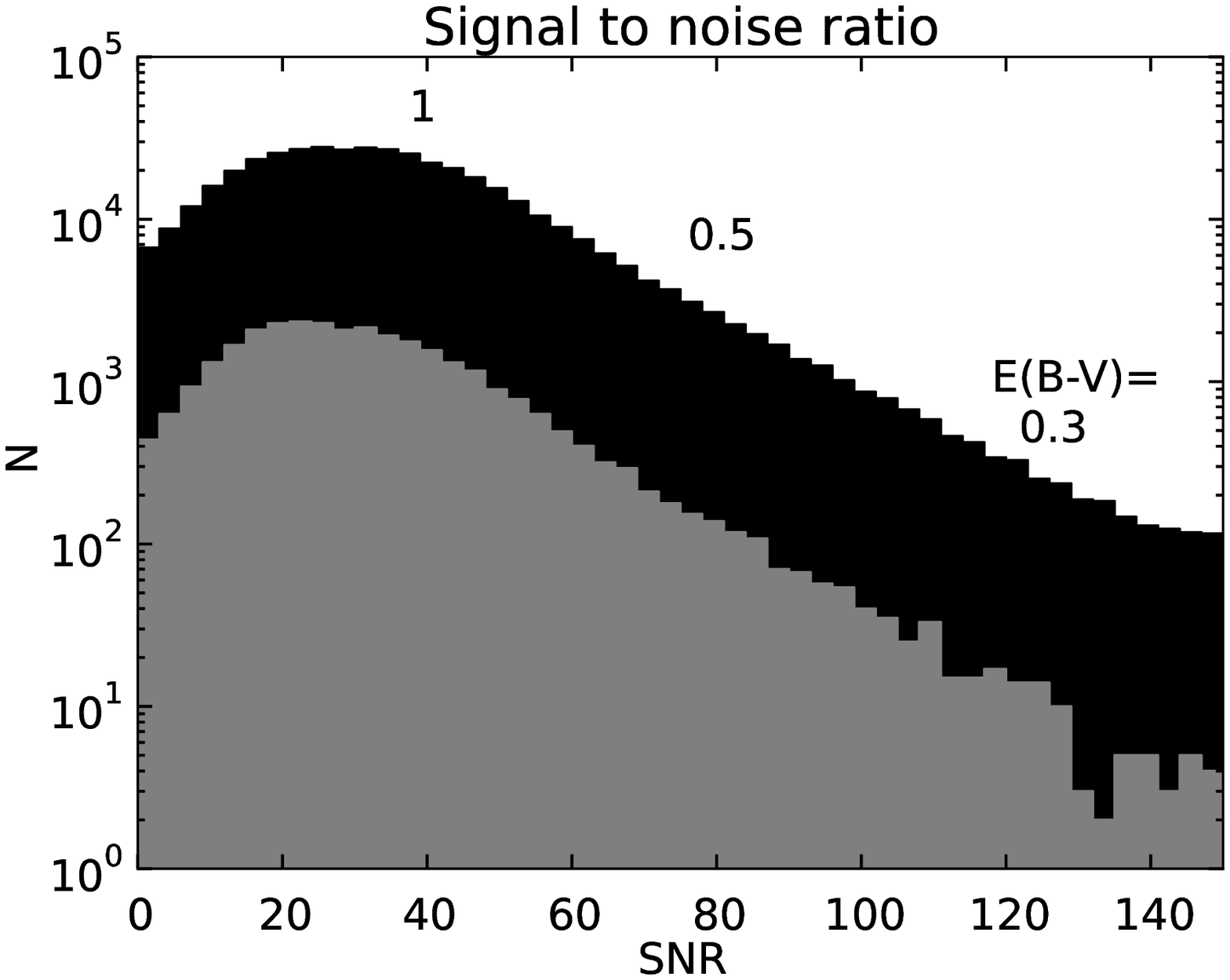} \includegraphics[width=\columnwidth]{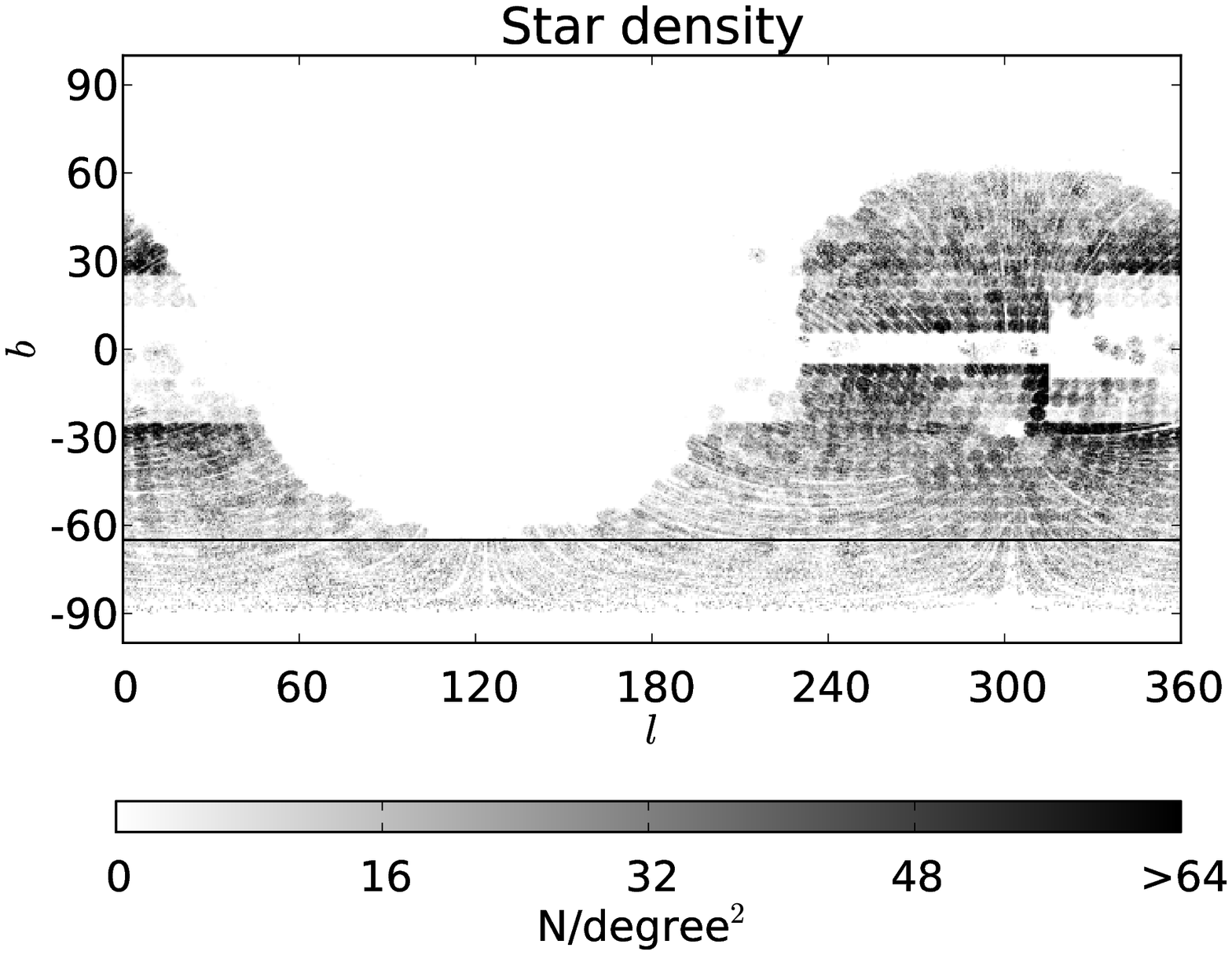}\\
\includegraphics[width=\columnwidth]{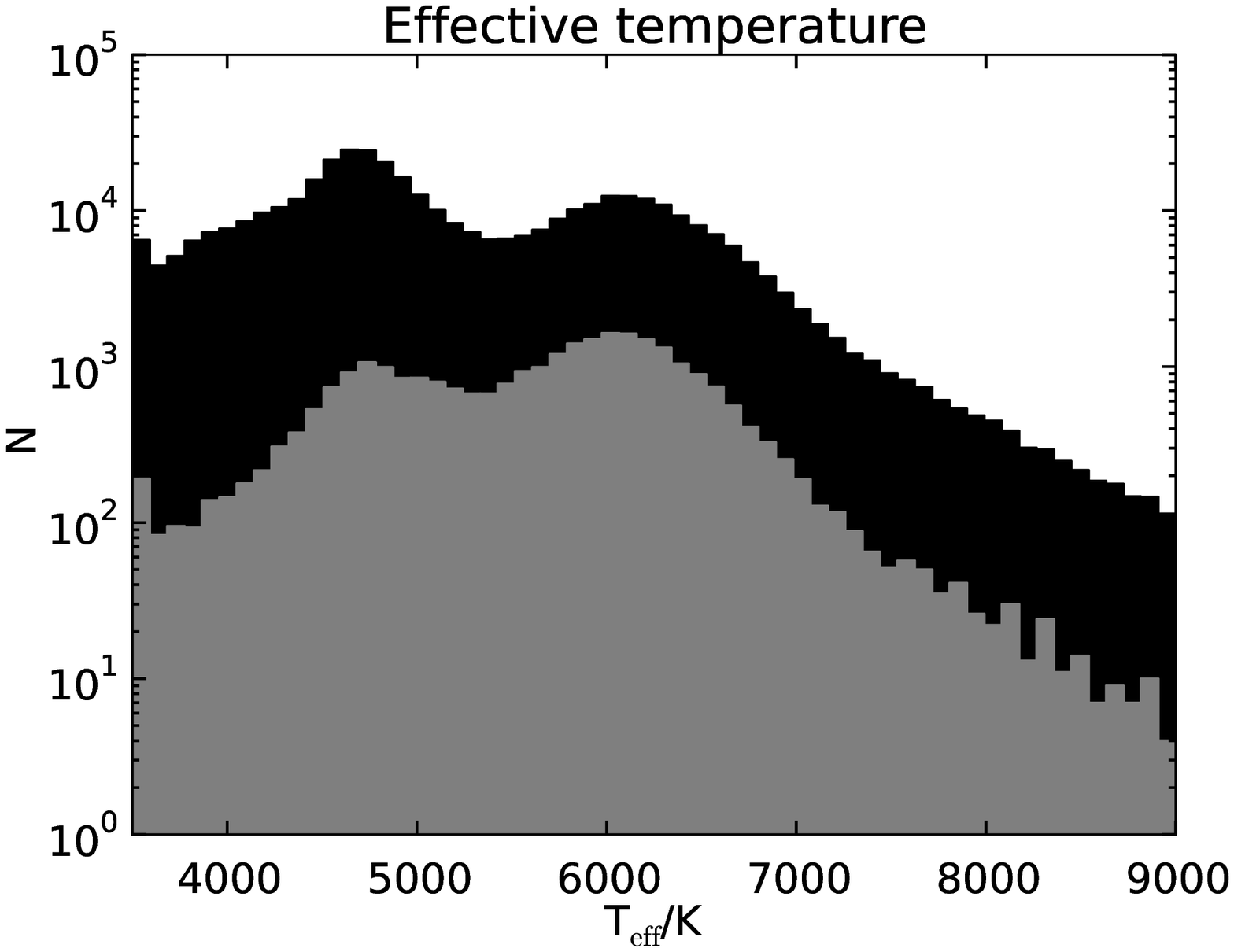} \includegraphics[width=\columnwidth]{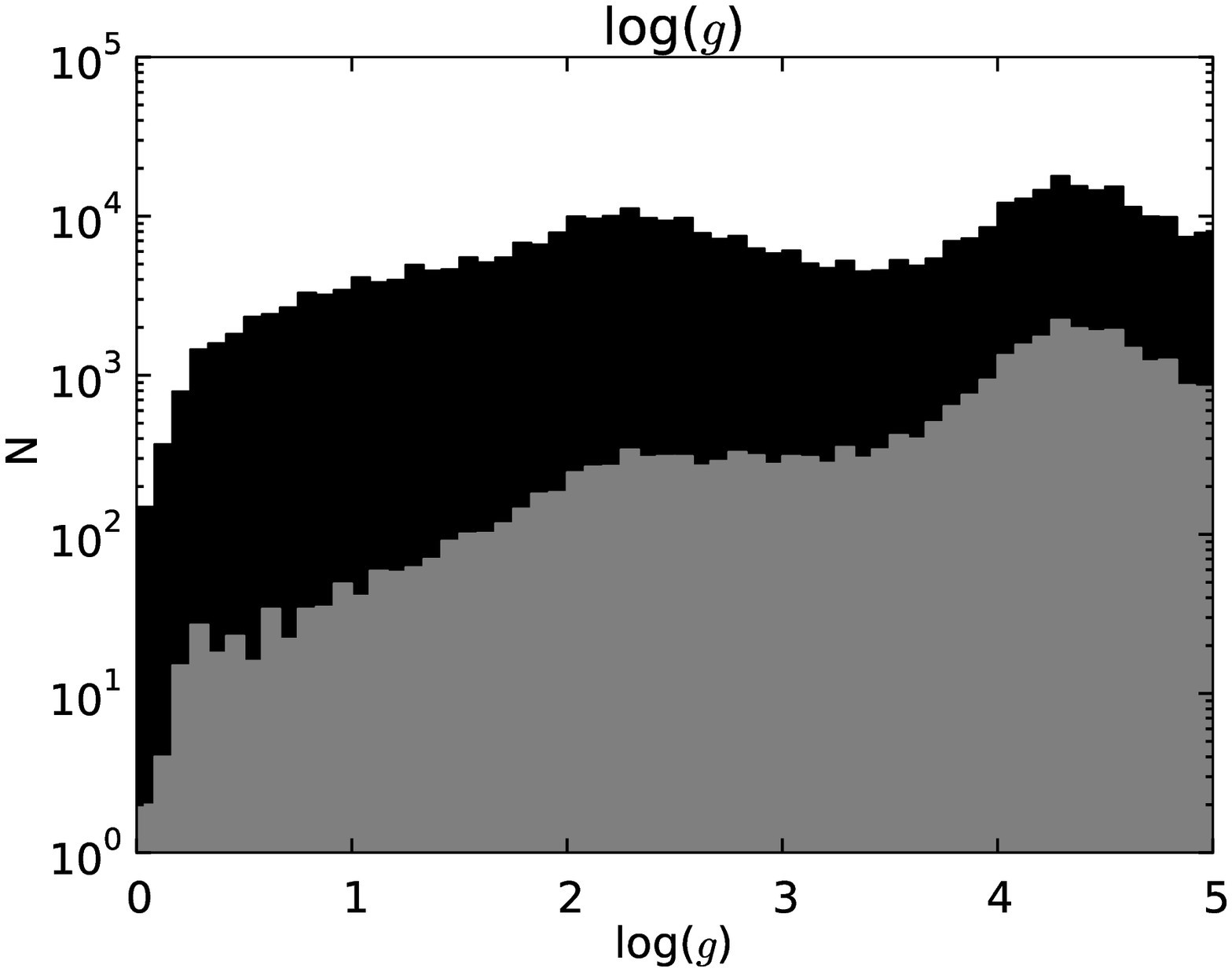}
\caption{\emph{Top-left: }Histogram of signal to noise ratio for all stars (black) and for stars with $b<-65^\circ$ (gray). Stars with S/N more than 150 are not shown. The numbers on top of the histogram show the reddening $E(B-V)$ that is required for the DIB to have the central depth equal to 1 standard deviation of the noise at that S/N (a relation from \citet{munari08} was used for this estimate). \emph{Top-right: } Distribution of stars on the sky. $b=-65^\circ$ is shown with a line. \emph{Bottom-left: } Effective temperature distribution for all stars (black) and for stars with $b<-65^\circ$ (gray). Stars with temperature above 9000 K are not shown. \emph{Bottom-right: } Histogram of gravity distribution for all stars (black) and for stars with $b<65^\circ$ (gray).\\}
\label{fig:quality}
\end{figure*}

Figure \ref{fig:quality} shows the range of several properties of the stars and their spectra. The observed fields cover declinations below zero, but avoid the Galactic plane at $-5^\circ<b<5^\circ$ and most of the Galactic bulge in the region with  $-25^\circ<b<25^\circ$  and $45^\circ>l>315^\circ$. This means that there are almost no regions with high interstellar extinction covered and all observed DIBs will be weak. In order to detect a weak DIB in a given spectrum, it must have a high S/N ratio. To detect the DIB 8620 in spectra of RAVE stars directly, the reddening of the sightline must be above 0.2 for the vast majority of stars and close to 1 for an average star. There are only a few observed regions with such a high reddening. 

The procedure to model the stellar spectrum described in this paper requires a large number of observed stars with no or very low reddening. Such stars were selected from the fields with $b<-65^\circ$, which include 7.5~\% of all observed stars, 36,308 in total. As the stars are less abundant with rising temperature, the method becomes unreliable for stars with $T_{eff}>7000 \mathrm{K}$, because there are not enough stars at $b<-65^\circ$ to construct a stellar template at this temperature. Because the shape of the spectrum is less dependent on $\log(g)$, the method works at all gravities and becomes unreliable only at very extreme values of $\log(g)<0.5$. The same holds for the metallicity. 

\subsection{Dividing by stellar spectrum}
Because the DIB feature is expected to be very weak in most of the spectra, a precise elimination of the stellar component is the most important step in the analysis. Usually synthetic spectra are used for the stellar component, but they have many disadvantages. They are calculated on a grid of parameters, most often with a step of 250 K to 1000 K for cool and hot stars, respectively. Other parameters are sampled at similar relative steps. Such a grid proves to be too widely spaced and some weak stellar lines suffer from inappropriate oscillator strengths. Divided spectra show significant artefacts if synthetic templates are used, mostly close to the lines inflection points or lines centers. Fe \textsc{i} line at 8621 {\AA}, the strongest line that overlaps with the DIB, is no exception. There are also some spectral lines missing in synthetic spectra, but not in the band around the DIB, so this does not affect the measurements of the DIB profile directly. For normal stars the divided spectrum shows artefacts as large as 4~\% of the continuum. DIB 8620 is expected to be shallower than this. 

\begin{figure*}[!p]
\centering
\includegraphics[width=18cm]{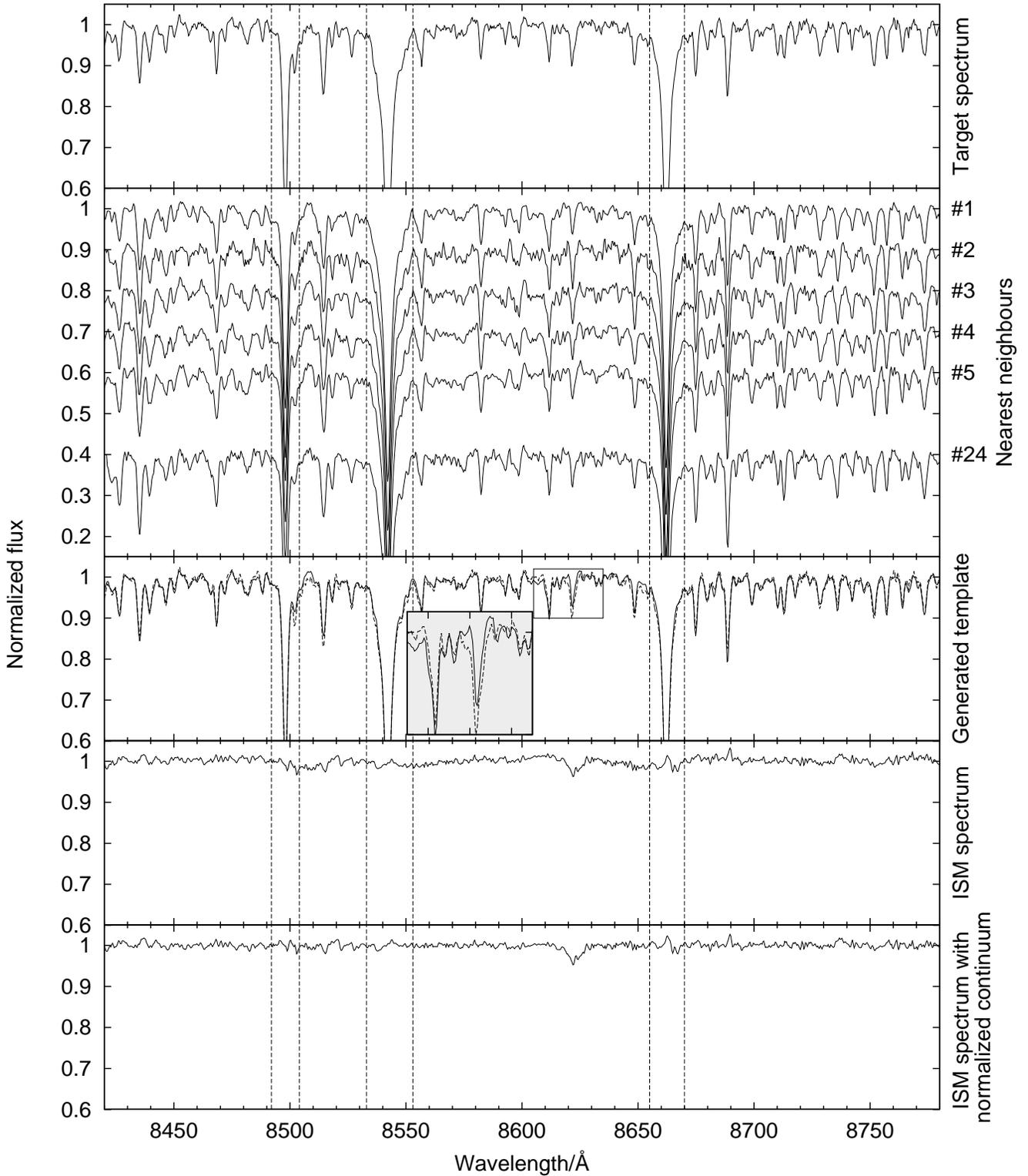}
\caption{Illustration of the main steps of the procedure to determine the ISM contribution to the spectrum of the star HD 149349. The spectrum has a S/N=150, parameters of the star are T$_{eff}$=4900~K, log$(g)$=2.0, and metallicity=-0.54. From top to bottom the following spectra are shown: The target spectrum, which is the spectrum from which we want to extract the DIB. Second pannel shows similar spectra found among the unreddened stars. The 24 best matches are taken in this case (of which the best 5 and the 24th spectrum are shown in the second plot). They are averaged into a template for the stellar component of the target spectrum (third plot). \textbf{The target spectrum is plotted with a dashed line. The marked region around the DIB is magnified for further clarity.} The target spectrum is then divided by this \textbf{template spectrum} to get the ISM spectrum in the fourth plot. Because the continuum normalization was slightly different for the target spectrum and the spectra used for the template, this difference shows itself in the ISM spectrum. A continuum is fitted again to get the final spectrum in the bottom plot. The dashed vertical lines mark the areas of the Ca~\textsc{ii} lines, where the weights for calculating nearest neighbours were lowered (see text). }
\label{fig:procedure}
\end{figure*}

Around half a million observed stars make a database large enough to include many spectra of different stars with very similar parameters, much closer to each other than the grid size in the database of the synthetic spectra. Many stars have high Galactic latitudes, where only a negligible amount of DIB component is expected. We assume that their spectra are free of interstellar, or any other, but stellar component. The expected equivalent width of the 8620 DIB can be estimated from the color excess from \citet{schlegel98}. \citet{schlegel98} give color excesses for objects outside the Galaxy, so the given reddening is only the upper limit for reddening for stars in our galaxy. For $b<-65^\circ$ the mean E(B-V) is 0.016. This corresponds to an equivalent width of 5.8 m\AA\ or a depth of around 0.08~\% of the continuum, if a FWHM of 5.59~{\AA} is assumed \citep{jenniskens94} and if average relation between reddening and DIB's equivalent width from this paper holds. This is the maximum error we expect when using spectra at high Galactic latitudes for the most inconvenient case. Because the \citet{schlegel98} values are known to overestimate the reddening \citep{yasuda07, kohyama13} and because most stars have lower reddening than the calculated limit, the real errors are much lower than that. Several spectra are averaged to produce the stellar spectrum with higher S/N ratio, so any stars with possible peculiar high reddening are averaged  out.

The procedure described in the following paragraphs is illustrated in Figure \ref{fig:procedure}. The spectrum used in this illustration is one of the few high S/N spectra that show the DIB and is therefore an unusual case, but very convenient for the illustration.

For each spectrum from which we want to extract the ISM spectrum, we find several spectra at $b<-65^\circ$ that most closely resemble the original spectrum. These spectra are called nearest neighbours, as they lie closest to the original spectrum in the parameter space. 

The first step is to limit the number of spectra based on their stellar parameters. Only spectra with a derived effective temperature in the range of $\pm20\ \%$, $\mathrm{log}(g)$ in the range of $\pm$0.6~dex, metallicity in the range of $\pm$0.4~dex, and $v\mathrm{sin}(i)$ in the range of $\pm40\ \%$ around the calculated values are retained. These limits are taken very liberally, as the errors of the calculated stellar parameters are always much smaller. In this way the selection of spectra is limited to around 500 stars. The next step is to compare the target spectrum to these 500 stars and to find the most similar spectra. Spectra are shifted into the same velocity frame and interpolated to have the same sampling. The estimator for the similarity of the spectra is the sum of the positive differences of all pixels. However, several areas must be weighted differently. The area around the DIB \textbf{(between 8612~{\AA} and 8628~{\AA})} is excluded from the difference calculations and the weights on the central part of all three {Ca~\textsc{ii}} lines are put to 30~\%. Ca~\textsc{ii} lines might show chromospheric activity \citep{zerjal13}. Such stars are most common among cool dwarfs \citep{matijevic12}. The differences between the spectra are highly dependent on {Ca~\textsc{ii}} lines, because they are the strongest lines in the observed spectral region. But also the chromospherically active stars can be used to model the stellar component of the spectrum. Therefore the influence of the Ca lines is intentionally reduced. The inclusion of active stars with Ca emission shows itself in the final spectrum as artifacts at the wavelengths corresponding to Ca~\textsc{ii} lines, but does  not influence the band around the DIB. \textbf{The whole observed spectral range is compared to find the nearest neighbours. Apart the Ca\textsc{II} lines, the spectra are dominated by Fe lines and a good match of the whole spectrum also means a good match of the DIB's neighbourhood, at it is also dominated by Fe lines.}

When all 500 spectra are compared with the original spectrum, up to 25 best matching spectra are averaged (with weights corresponding linearly to their S/N) into one stellar spectrum. For stars with more unusual parameters the number of spectra is smaller than 500. The smallest number of spectra that we ever used was 25. In this case only the 8 best matches are averaged into the final stellar spectrum. The target spectrum is then divided by this spectrum. We will call the divided spectrum the ISM spectrum. \textbf{The generated stellar spectrum is calculated in the same way, regardless of the S/N of the target spectrum. Because we are aiming for the stellar spectrum to have high S/N, the target spectrum has almost always lower S/N than it's generated stellar spectrum.}

\subsection{Combining spectra}

The mode of the S/N ratio distribution is around 25. With a few exceptions, this is too low to detect the DIB in an individual star. Spectra of stars in an arbitrary volume can be averaged in order to achieve a better S/N. Before the averaging, the spectra are shifted back into the local standard of rest. If there are any spectra left that do not have the stellar component properly eliminated, this process smooths out such deviations. The measured DIB equivalent width is then proportional to the averaged gas density of the DIB's carrier toward each star in the region. For example, in a volume with average E(B-V)=0.2, a spectrum with S/N around 300 is required for the DIB to have a depth equal to 3 standard deviations of the noise. With a S/N distribution with mode at 25, $\sim$100 spectra, occupying a certain volume, have to  be averaged to achieve such a high S/N. In reality the number is somewhat lower, because the weighted average is calculated with weights linearly depending on the S/N of the ISM spectrum.

To obtain the results presented in this paper, an average of 48 spectra was used. This is also the limiting factor in the spatial resolution due to the limited star density of the survey. \textbf{The star density in the observed volume varies significantly and so does the limiting spatial resolution.}

\subsection{Determination of DIB parameters}

The FWHM of DIB 8620 is around 5~{\AA}, which is more than the resolution of the RAVE spectra (around 1.15~{\AA} at the wavelength of 8620). This range is covered by around 12 pixels. \textbf{\citet{jenniskens94} actually report 2 features, described by two Gaussians with FWHMs of 1.86 and 5.59~{\AA} at almost the same position, the narrower one being about three times weaker than the wider one. However, because the blend of both features is not far off from a Gaussian shape and because fitting two Gaussians to low resolution RAVE data would be hard to automate and do it precise, we used only one Gaussian to represent the DIB profile.}

The equivalent width of the DIB is measured in two ways. The first one is by using the definition (integration):
\begin{equation}
W_e=\int_{\lambda_1}^{\lambda_2}\left( \frac{F_c-F_\lambda}{F_c}\right)d\lambda,
\end{equation}
where $F_c$ and $F_\lambda$ are fluxes of the continuum and spectrum, respectively. The error is estimated from the noise in the regions near the DIB. The second method consists of fitting a Gaussian profile to the DIB using the Levenberg-Marquardt method. This method gives position, width and equivalent width of the fitted profile, as well as the errors from the covariance matrix. The resolution and S/N are too low to fit more than just a simple Gaussian profile.

The limited S/N is not the only source of error on the equivalent width. A large error arises from the uncertainty of the continuum determination. Even tough the used spectra were normalized, the ISM spectrum did not have a uniform continuum. The deviations from a constant continuum were of the order of 1~\%. Legendre polynomial continuums of different orders (up to 35th order) were fitted to the whole ISM spectrum (around 800 data-points), producing a series of spectra with different continuum normalization. The above procedure of parameters determination was performed on each of the spectra in the series and the scatter in each parameter was added to the measured errors. The average error of the $W_e$ after this operation is around 10~\%, which is normal for such spectra and spectral lines. \textbf{Not normalizing the ISM spectrum prior the profile fitting would require an additional free parameter (the level of the continuum) to be introduced and would result in less stable calculation of the profile fit.}

\begin{figure}
\centering
\includegraphics[width=\columnwidth]{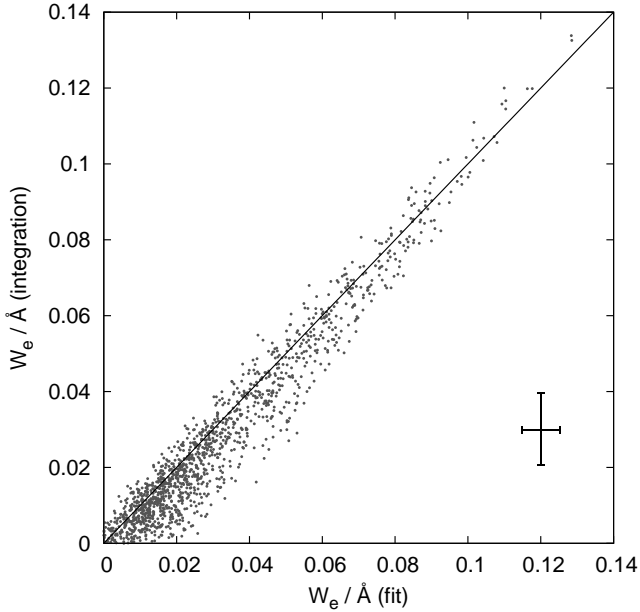}
\caption{Correlation between fitted equivalent width and integrated equivalent width. The 1:1 correspondence is marked with a line. The typical error for each datapoint is marked with a cross. Datapoints are the same as used for Figures \ref{fig:schlegel} and \ref{fig:binney}.}
\label{fig:corr}
\end{figure}

The fitting of the DIB profile is completely automated, hence there may be the concern, that in the low S/N spectra with weak DIB, the method might overestimate the equivalent width of the DIB. Several constraints are given for the DIB profile, one of them being that the DIB is an absorption feature. The DIB only covers around 12 pixels, so it is likely that in some cases the noise will resemble the DIB. Because the position of the DIB is also fitted, the procedure will search for such a dip in the area where DIB is expected to be. A simple integration over the same area is less sensitive to this effect, because coincidental rises in the continuum are equally probable and average out the dips. Figure \ref{fig:corr} shows the extent of this problem. A typical overestimation at $W_e=0$ is 3.5 {m\AA} and gradually drops to zero at larger equivalent widths. Data described in Section \ref{sec:corr} was used to make this comparison. 

\section{Tests}
\label{sec:tests}

\subsection{Neighbours search}

A nice test of the nearest neighbour method is to compare the stellar parameters and to classify the individual stars and their nearest neighbours. 

\begin{figure*}
\centering
\includegraphics[width=\columnwidth]{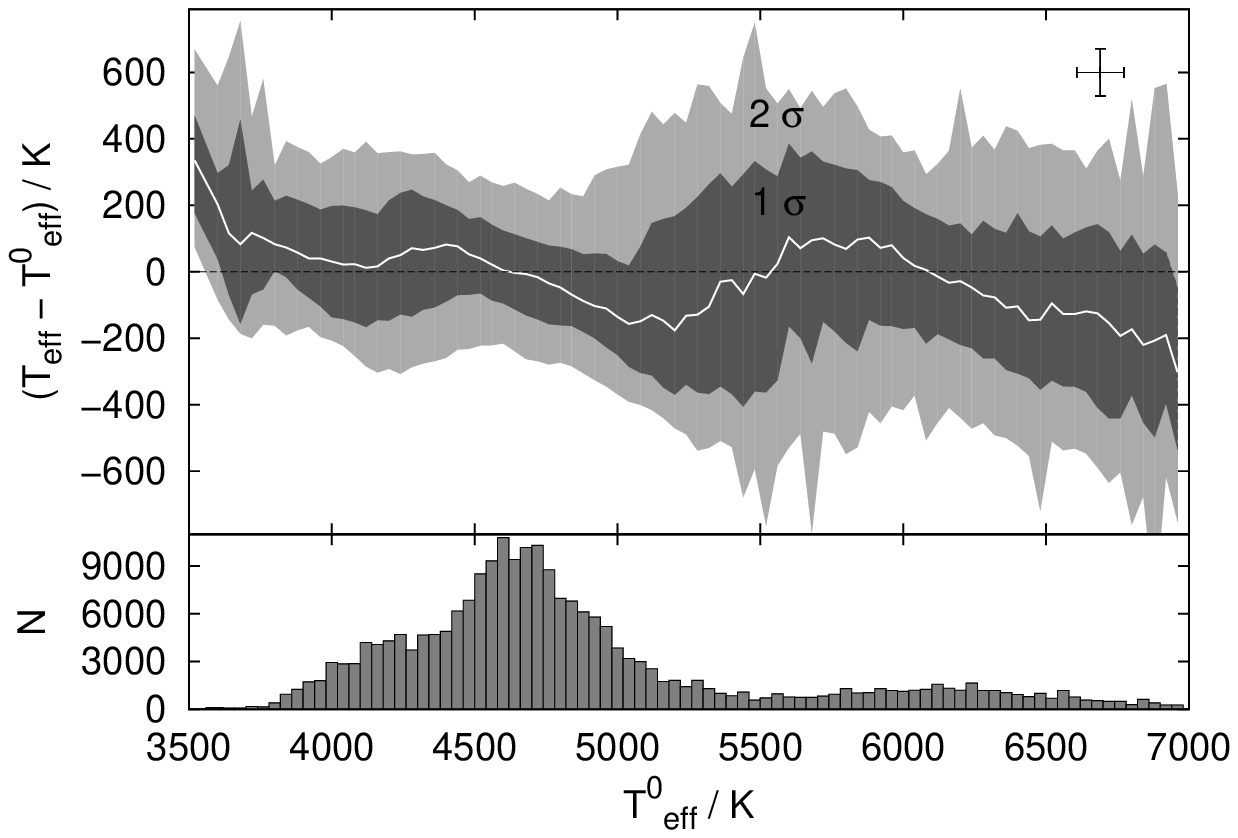} \includegraphics[width=\columnwidth]{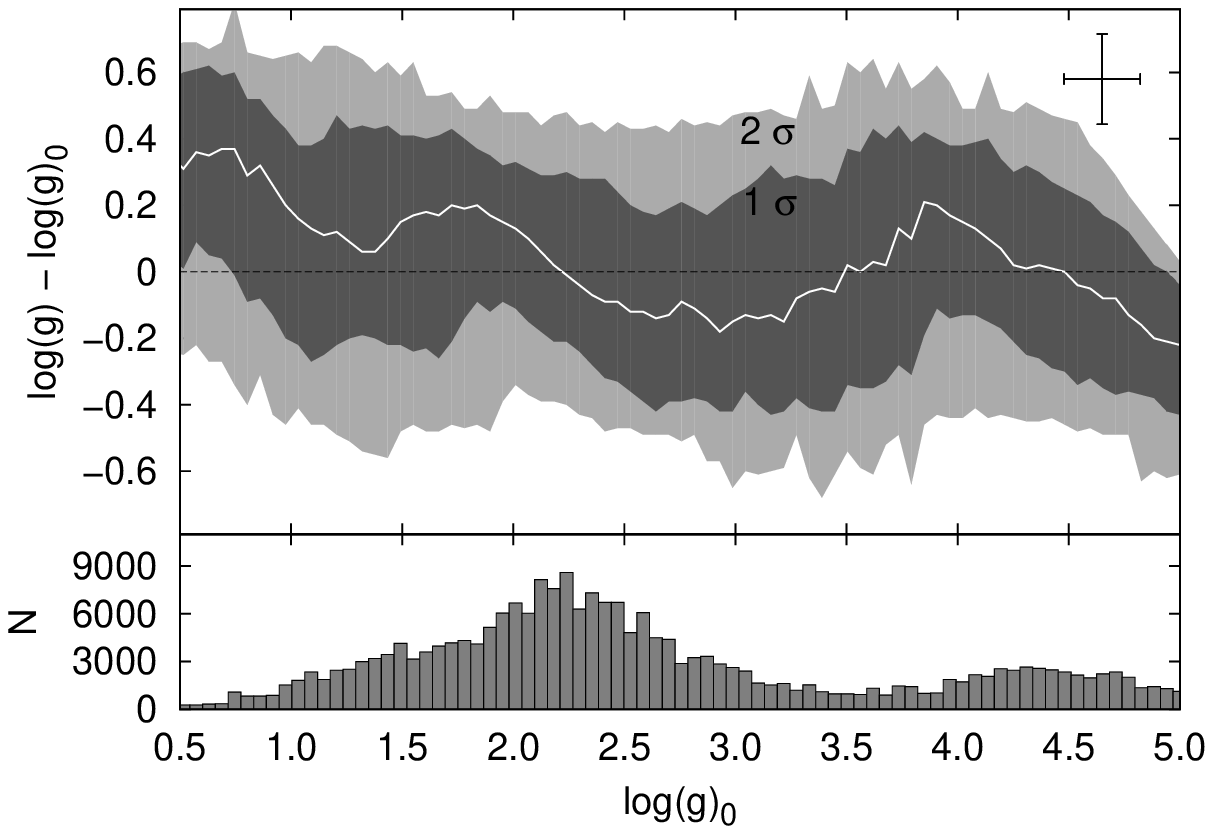}\\
\includegraphics[width=\columnwidth]{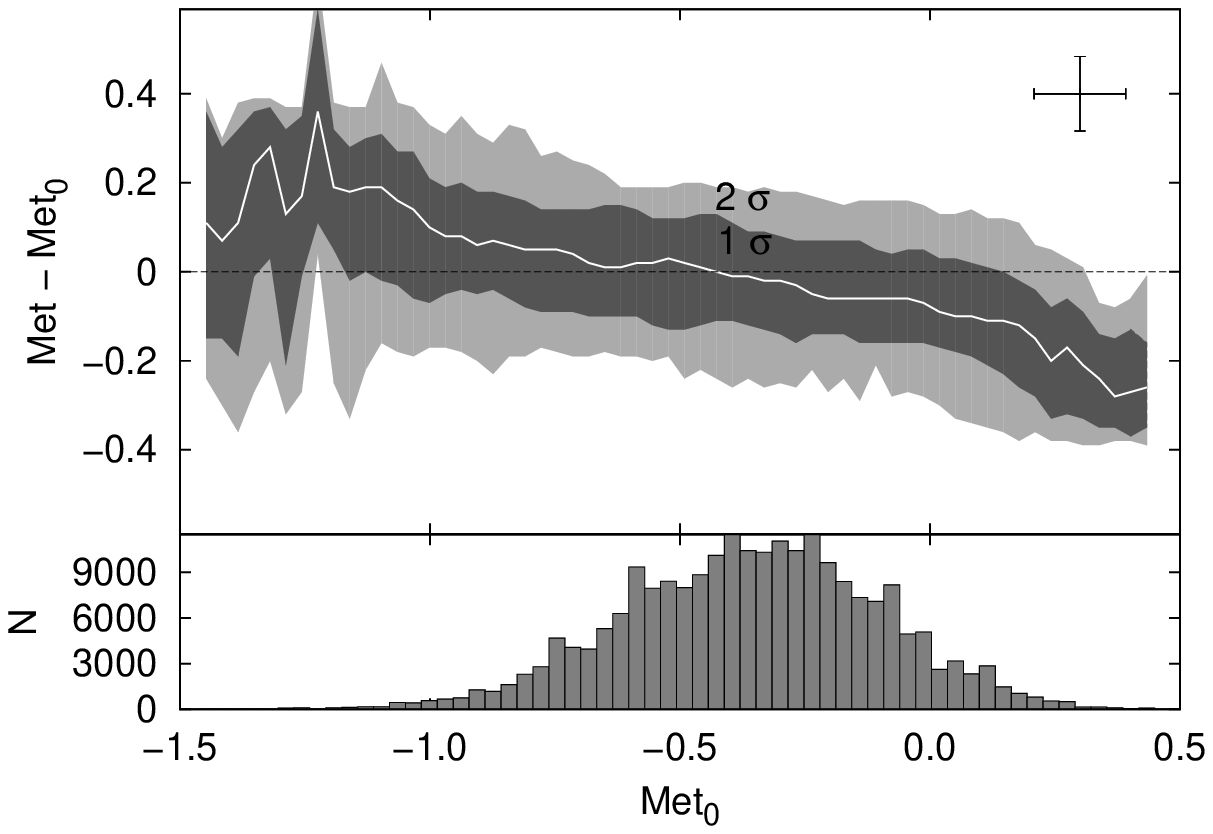}
\includegraphics[width=\columnwidth]{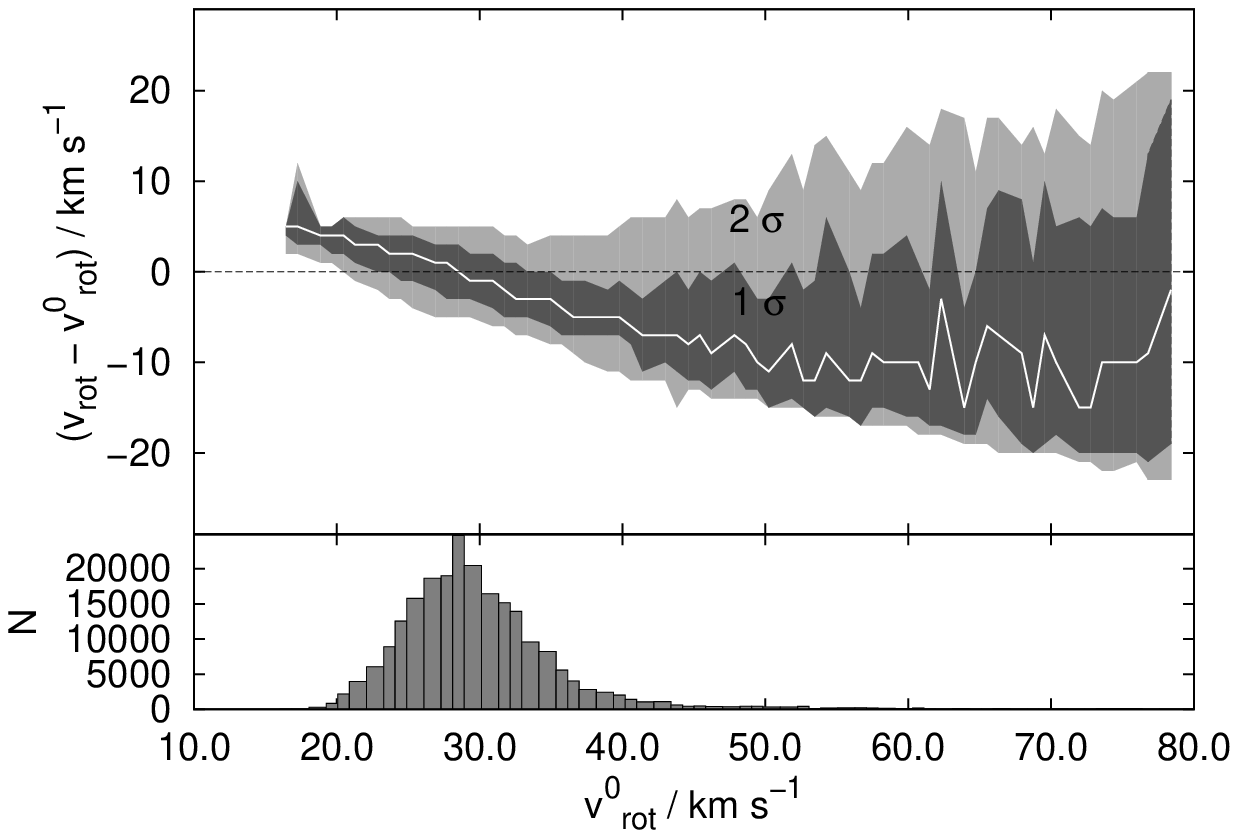}
\caption{Differences between the parameters of nearest neighbours and parameters of individual stars. Parameters compared are effective temperature (top left), gravity (top right), metallicity (bottom left) and projected rotational velocity (bottom right). The black line shows the median of the distribution of the neighbours' parameters. One and two standard deviations of the distribution are marked in dark and light gray, respectively. The typical error of the derived parameters is shown in the top right corner (except for the rotational velocity, where no error was calculated). All plots are accompanied by a histogram of the number of stars used for the statistics. For each star up to 25 neighbours were found. Keep in mind that the value of the rotational velocity is highly influenced by the spectrograph resolution.\\}
\label{fig:test_comp}
\end{figure*}

The method of classification is described in \citet{matijevic12} and was used to classify most of the stars in DR4, used in this paper. Most of the stars (82~\%) are classified as normal stars, around 7~\% of the stars are unclassified for different reasons or have nearest neighbours with different classification (these stars will be named unclassified in this paper), 4~\% of stars are classified as stars with emission in the Ca~\textsc{ii} triplet, 3~\% are spectra with problems with continuum fitting or wavelength calibration and the remaining 4~\% are classified as binary stars, TiO band stars, hot (T$>$7000 K) stars, cool (T$<$3500 K) stars, peculiar stars and carbon stars (in order from the most frequent to the least frequent classes). 

Whether the individual stars share the same classification with their nearest neighbours is a powerful test of our method. For normal stars on average 98.7~\% of the neighbours are normal stars as well. The remaining neighbours have mostly unclassified spectra. The opposite holds for the unclassified spectra, where most of the neighbours are normal stars. Ca~\textsc{ii} emission spectra have nearest neighbours that are in the same class only in 55~\% of the cases. Because most of these stars have weak emission and because different weights are set for the Ca~\textsc{ii} triplet, the normal stars (representing other 45~\% of the neighbours) are good matches to the spectrum with weak emission. Spectra with TiO band have nearest neighbours that belong into the same class in 17~\% of the cases. However these stars are so rare that there are not enough stars of the same type in the DR4 and if we force the method to find a certain number of neighbours, it will usually include some of the normal stars. For other classes the comparison  is difficult due to an even lower number of stars in each class.

This test gave positive results, because the stars have nearest neighbours of the same class much more often than pure coincidence would imply.

\begin{figure}
\centering
\includegraphics[width=\columnwidth]{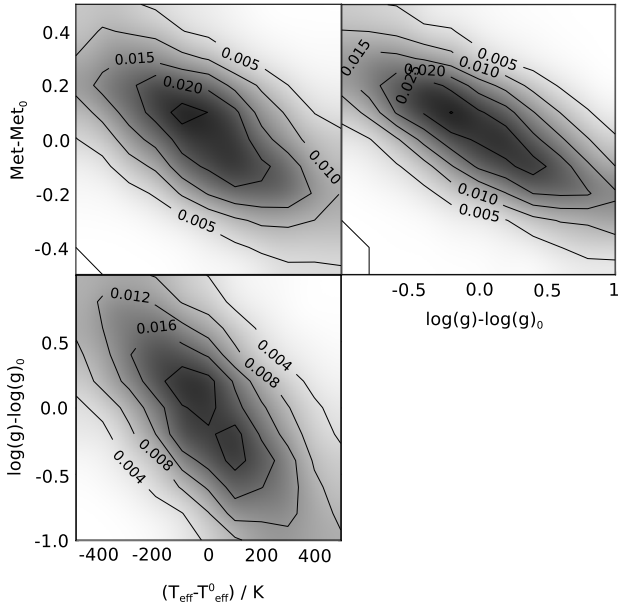} 
\caption{Correlations between the deviations of three main stellar parameters.}
\label{fig:tect_cor}
\end{figure}

The second test was to compare the derived stellar parameters of individual stars and their nearest neighbours. Temperature, gravity, metallicity and projected rotational velocities were compared. Figure \ref{fig:test_comp} sums up the results. 

The width of the distribution is a combined effect of the final number of spectra from which the neighbours are selected and the accuracy of the stellar parameters that is achievable at a certain S/N. Temperature is the leading parameter that drives the shape of the spectrum. Therefore it is calculated most precisely. A typical relative error is around 5~\%. In this case the width of the distribution is mostly affected by the final number of spectra in the database. In order to construct a good stellar spectrum template, several spectra must be averaged. If there are not enough spectra with exactly the same temperature (within the error), spectra with similar temperature are taken. Gravity is calculated with lesser accuracy and the width is mostly governed by noise. This is even more so in the case of metallicity, where the width of the distribution matches the accuracy of the calculated metallicity. We also compared the projected rotational velocity. Special care is needed in the interpretation of the results, as the rotational velocity at these noise levels is almost indistinguishable from the effect of the changing resolution of the spectrograph. Because we are only interested in finding the best matching spectra, it is not important which of the two parameters the $v_r \sin{(i)}$ really traces.

On average all the distributions are centred at 0. However there are non-negligible deviations in the distribution at a varying value of the parameter. Because the spectra have low S/N, a small perturbation in the spectrum can be attributed to a perturbation in more than one parameter. The correlation between the parameters is plotted in Figure \ref{fig:tect_cor}. This does not mean the nearest neighbours are not good matches, it just shows that there is a certain degree of uncertainty in spectra with mode of the S/N distribution around 25 \citep{kordopatis11}. 

We conclude that the nearest spectra have good matching parameters and classification to their target stars.

\subsection{Effective temperature invariance test}

We want to check if the measured equivalent width of the DIB 8620 is independent of the temperature of the stars used to calculate the ISM spectrum. We took different regions on the sky where the DIB is expected to be strongest and calculated its equivalent width only with stellar spectra within a temperature range of 100~K. The regions were all located along $230^\circ<l<315^\circ$ and in two bands between $5^\circ<b<10^\circ$ and $-10^\circ<b<-5^\circ$. All regions covered $5^\circ \times5^\circ$ in $l$ and $b$ and a 0.5~kpc range in distance. We used regions with the nearest distance of 1~kpc, 1.5~kpc and 2~kpc. Together this makes 102 different regions. Only stars with a temperature range of 100 K were used each time the equivalent width was calculated. Equivalent widths were then normalized to the average equivalent width in each region, so regions with different density of DIB carrier can be compared. 

\begin{figure}
\centering
\includegraphics[width=\columnwidth]{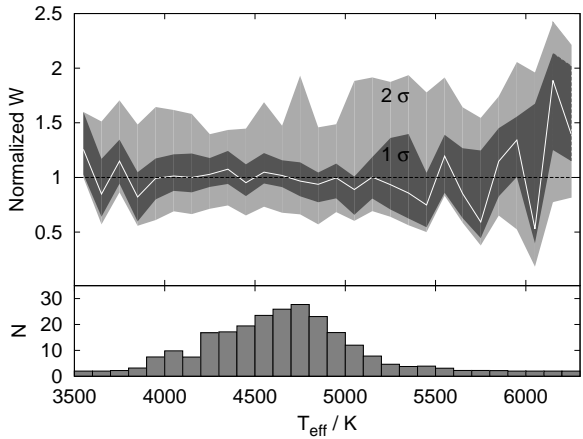}
\caption{\emph{Top: }Average normalized equivalent width for different temperatures of used stars. One and two standard deviations of the distribution of 108 measurements for every temperature class around the average value are plotted in dark and light gray. Regions with less than 2 stars were excluded from the calculations, so the distribution consists of less than 108 measurements for the least abundant temperature classes. \emph{Bottom: }Average number of stars in the selected region for each temperature class.}
\label{fig:test_temp}
\end{figure}

Figure \ref{fig:test_temp} shows the result of the analysis. The average value of the normalized equivalent width in all regions is plotted against the temperature of the stars. The result is consistent with 1 for all temperatures, so the calculated equivalent widths are independent from the temperature of the stars used in the procedure. For a precise measurement of the DIB profile, we had to use regions as close to the Galactic plane as possible. Because the stars were divided even further into the temperature classes, only the regions with highest star density could be used, as the spectra of several stars must be combined to gain a S/N high enough to detect a DIB. This leads to a small number of regions considered in the calculation and to a wide distribution of the measurements along the $W=1$ line.

\section{Correlation with reddening}
\label{sec:corr}

The correlation of DIB 8620 equivalent width with reddening was previously studied in \citet{munari08,munari00}, \textbf{\citet{wallerstein07}} and \citet{sanner78}. \citet{munari08} studied the DIB in 68 RAVE spectra of hot stars. Local stellar spectrum was approximated by a 6th order Lagrange polynomial and the spectrum was integrated over the range of the DIB to get the equivalent width. The relation from \citet{munari00} is obtained from spectra of 37 hot stars, observed with Asiago Echelle spectrograph. \citet{sanner78} only give measured equivalent widths and the color excess of 15 hot stars observed with Coud\'e spectrograph of the 2.7 meter McDonald Observatory telescope, so the parameters of the linear relation were calculated by us. \textbf{\citet{wallerstein07} calculated correlation between the reddening and the equivalent width of the DIB 8620 from a sample of 53 stars, but \citet{munari08} dispute their results.}

\subsection{Correlation for hot stars}

In order to have another correlation with which to compare the results of the new method, we repeated the measurements done in \citet{munari08}. We detected the DIB in spectra of 144 stars with an effective temperature larger than 8000 K. We fitted a Gaussian profile to the DIB, together with a second order polynomial, which represents a local continuum. \textbf{Only the region between 8605~{\AA} and 8660~{\AA}, an area between the 14th Paschen hydrogen line and a blend of Ca\textsc{II} and 13th Paschen hydrogen line was taken into the account when fitting the Gaussian profile and the local continuum. A narrow N\textsc{I} line at 8630~{\AA} was also omitted.}

\begin{figure}
\centering
\includegraphics[width=\columnwidth]{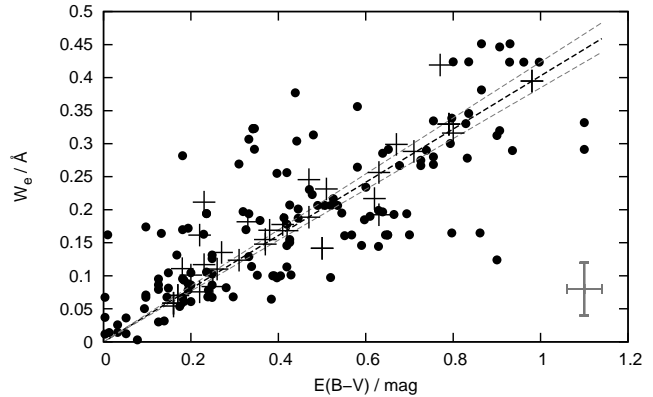}
\caption{Equivalent width  of the DIB 8620 compared to the reddening from the literature. 1 $\sigma$ lines are plotted in gray. Bottom-right errorbars show typical errors for the measured points. Crosses mark 31 stars from \citet{munari08} that we were able to fit as well. For these stars E(B-V) values from the same paper were used.}
\label{fig:umeritev}
\end{figure}

The resolving power of the RAVE spectra is around 7500. This is high enough to resolve a broad DIB feature, but low enough to discard any substructure or peculiarities in the DIB profile when fitting a simple Gaussian curve to it. Spectra of hot stars are smooth enough, with no distinctive features around the DIB to make the elimination of the stellar component unnecessary. There is also no contamination by telluric lines. Measurements of the DIB were compared with the reddening from the literature (\citep{neckel80}  -- 18 stars, \citep{savage85} -- 6 stars, \citep{guarinos92} -- 32 stars, \citep{bailer-jones11} -- 82 stars, \citep{gudennavar12} -- 3 stars). Some stars are present in more than one catalogue. In addition, we used stars from \cite{munari08} and used reddenings from the same paper. However we were only able to detect the DIB in 31 spectra out of 68. In other spectra from \citet{munari08} the measured DIB's equivalent width was consistent with zero on a one standard deviation level due to high noise. The measurements are shown in Figure \ref{fig:umeritev}. A linear relation was fitted to the measurements, taking the measurement errors into account.

All the coefficients $a$ for the linear relation ${E(B-V)=a\cdot W_e}$ between reddening and the DIB's equivalent width are listed in Table \ref{tab:redd}. Measurements based on a sample of hot stars have all been carried out with different detectors and/or with different methods to measure the equivalent width.  \citet{sanner78} observed with a reticon detector and integrated the spectrum to get the equivalent width, \citet{munari00} observed with a thick CCD detector and measured the equivalent width by integration, \citet{munari08} used RAVE observations, made with thinned CCD detector, the same as in this paper, except that we fitted the continuum differently and also fitted the DIB, instead of integrating it. \textbf{Results from \citet{wallerstein07} are inconclusive as discussed in \citet{munari08}.}

\begin{table}
\caption{Coefficient of the linear relation between DIB 8620 and reddening as given in the literature, compared to the calculations from this paper.}
\begin{tabular}{l c c}
\hline\hline
source & coefficient $[ \mathrm{mag / \AA} ]$ & std. deviation\\\hline
\citet{munari00} & 2.69 & 0.03\\
\citet{munari08} & 2.72 & 0.03\\
\citet{sanner78} & 2.85 & 0.11\\
\citet{wallerstein07} & 4.61 & 0.56\\
this paper$^a$ & 2.48 & 0.12\\
this paper$^b$ & 2.40 & 0.12\\
this paper$^c$ &\ 2.49$^\star$ & 0.23\\\hline
\multicolumn{3}{l}{$^a$ hot stars}\\
\multicolumn{3}{l}{$^b$ \citet{schlegel98} maps}\\
\multicolumn{3}{l}{$^c$ Bayesian reddening}\\
\multicolumn{3}{l}{$^\star$ and offset of (0.028 $\pm$ 0.002) mag}
\end{tabular}
\label{tab:redd}
\end{table}

\subsection{Correlations for cool stars}

For correlations calculated with cool stars we were not able to use individual stars. High S/N (S/N of more than 300 is preffered) cool stars that have high redddening are not abundant in the RAVE survey and in addition the published reddenings in the literature matched only a few stars from DR4.

Instead, we combined several spectra in a small region to get one averaged high S/N spectrum where we can measure the DIB profile with high precision, even if the reddening is small. For the densest regions we used $2^\circ\times2^\circ$ wide areas in $l$ and $b$ that were 0.1 kpc deep in distance. For less populated regions we used areas $5^\circ\times5^\circ$ wide and 0.2 kpc deep. We did this for $-35^\circ<b<35^\circ$. The reddening of the region was calculated from the reddenings for all stars in the region, where we set the weights for averaging in the same way as the weights of the ISM spectra. 

The reddening for individual stars is calculated as a by-product of the distance calculations. A Bayesian distance finding algorithm is presented in \citet{burnett10}. To include the effects of the interstellar dust, several modifications were made \citep{binney13}. The input parameters are J, H, and K magnitudes from the 2MASS catalog \citep{skrutskie06}, stellar parameters from RAVE and color excesses from \citet{schlegel98}, corrected according to \citet{arce99}. In addition, several assumptions for the shape of the Galaxy \citep{juric08}, metallicity and age distribution \citep{haywood01, aumer09, carollo10} and isochrones \citep{bertelli08} are made.

\begin{figure}
\centering
\includegraphics[width=\columnwidth]{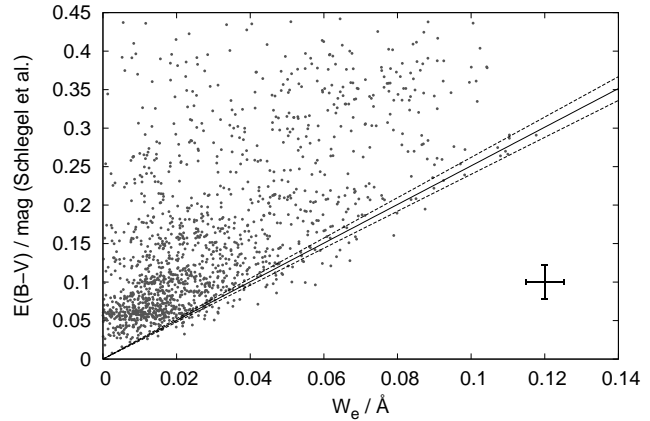}
\caption{Correlation between fitted equivalent widths for different regions and reddening from \citet{schlegel98}. The best fit of the lower envelope and one standard deviation are marked with the solid and dashed lines, respectively. The errorbars in the bottom right corner show typical errors for equivalent width and reddening. The reddening error is a combination of the errors for individual stars in a region and the scatter of the reddenings. The parameters of the fit are listed in Table \ref{tab:redd}. See text for the description of the fitting procedure.  \textbf{There are 421 volumes of size $5^\circ\times5^\circ\times0.2\ \mathrm{kpc}$ and 1745 volumes of size $2^\circ\times2^\circ\times0.1\ \mathrm{kpc}$ used for this plot.}}
\label{fig:schlegel}
\end{figure}

\begin{figure}
\centering
\includegraphics[width=\columnwidth]{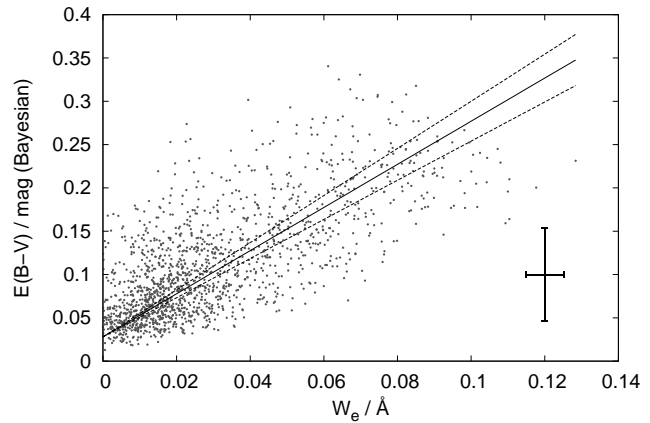}
\caption{Correlation between fitted equivalent widths for different regions and Bayesian reddening. The best fit and one standard deviation are marked with solid and dashed lines, respectively. Errorbars in the bottom right corner show typical errors for equivalent width and reddening. The reddening error is a combination of the errors for individual stars in a region and the scatter of the reddenings. The parameters of the fit are given in Table \ref{tab:redd}. \textbf{There are 421 volumes of size $5^\circ\times5^\circ\times0.2\ \mathrm{kpc}$ and 1745 volumes of size $2^\circ\times2^\circ\times0.1\ \mathrm{kpc}$ used for this plot.}}
\label{fig:binney}
\end{figure}

The correlation between the reddening E(B-V) and equivalent width of the DIB 8620 is represented in Figures \ref{fig:schlegel} (correlation with \citet{schlegel98}) and \ref{fig:binney} (correlation with Bayesian reddening). The parameters of the linear relation are given in Table \ref{tab:redd}.

\citet{schlegel98} do not give the reddening for individual objects. The product of \citet{schlegel98} is a map of extinction for the objects outside the Galaxy. For every object, or in our case every area on the sky, the given reddening is just the upper limit of the reddening toward the object. In addition, the reddening from \citet{schlegel98} is known to be overestimated \citep{arce99, yasuda07}, so the reddenings for the areas used in this study were corrected according to \citet{yasuda07}. With just the upper limit, the linear relation between the equivalent width and the reddening can not be acquired in the usual way, but the envelope must be fitted. 

\textbf{The difference between the \citet{schlegel98} and Bayesian reddening is evident from Figures \ref{fig:schlegel} and \ref{fig:binney}. If we compared the two reddenings for every data-point, the \citet{schlegel98} reddening would still represent the upper limit and would always be larger (within the error) than the Bayesian reddening.}

A condition used to fit a linear relation to \citet{schlegel98} data was to find the flattest line, where 32~\% of the points lying below the line have 1 $\sigma$ errorbars that do not extend above the line. Error lines were defined as linear relations, where the number of points under the line changes by 32~\%. No offset in the linear relation was considered. 

\begin{figure}
\centering
\includegraphics[width=\columnwidth]{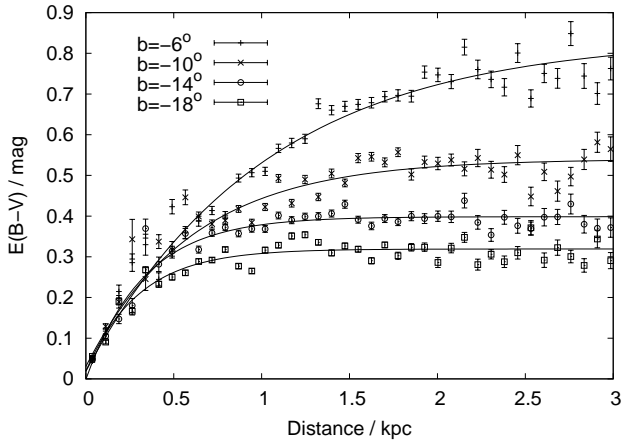}
\caption{Average Bayesian reddening of stars in regions at $b=-6^\circ$, $b=-10^\circ$, $b=-14^\circ$ and $b=-18^\circ$ (top to bottom) as a function of distance. An exponential model for dust density is fitted to each set of measurements. Only four examples of the 20 regions where this analysis was performed are plotted here. \textbf{Regions used to make this figure were selected differently than the regions used for Figures \ref{fig:schlegel} and \ref{fig:binney}, where the sightlines with $E(B-V)>0.3$ are mostly absent due to low star count in those volumes.}}
\label{fig:offset}
\end{figure}

When comparing the Bayesian reddening with measured equivalent widths, a strong offset of (0.028 $\pm$ 0.002) magnitudes is evident. The x-intercept is at -0.011 {\AA}, which is significantly more than -0.0058 {\AA}, which would be the x-intercept if the maximum error for the reddening of high Galactic latitude stars is assumed. No such offset is seen in Figure \ref{fig:schlegel}. There are also no points with reddening close to 0, which is unusual, as there are regions (like the ones in the local bubble) included, where the reddening is expected to be extremely low \citep{jones11}. Whether there is a systematic overestimation of Bayesian reddening can be tested by plotting reddening as it depends on distance. We used 20 regions, $2^\circ$ wide in $b$ and stretching from $l=230^\circ$ to $l=315^\circ$, and calculated the average reddening for stars at different distances. If we assume an exponential model for the dust density in $z$ direction, the function 
\begin{equation}
E(B-V)=A\left [ 1-\exp{(-d/d_0)} \right]+B
\end{equation}
can be fitted to any region. $d$ is the distance, $d_0$ is the distance where the dust density drops $e$ times (scale height divided by $\sin(b)$), and $A$ and $B$ are free parameters depending on the Galactic latitude $b$ and overestimation of the reddening, respectively.  Figure \ref{fig:offset} shows four examples of such an analysis. The average offset in all 20 regions was $(0.026 \pm 0.009)$~mag, which matches perfectly the offset of the linear fit in Figure \ref{fig:binney}. We conclude that the Bayesian reddening is overestimated by $(0.026 \pm 0.009)$~mag and that the systematic error for the equivalent width of the DIB arising from our assumption of unreddened stars at high Galactic latitudes is negligible.

\section{Conclusions and discussion}
\label{sec:conclusion}
We show in this paper that a simple idea of combining observed spectra into a stellar spectrum template can be developed into a powerful tool for the extraction of interstellar lines from a large number of spectra. This method can only be used on a large number of spectra, preferably observed with a single instrument. It is independent of stellar parameters, so it uses less assumptions than using synthetic templates and is not vulnerable to errors in the synthetic spectra, like missing lines, or different continuum normalization of synthetic and observed spectra. A known radial velocity of the observed star is desirable, but not necessary. If there is none the method still works, but requires more computational time to calculate the shift of each spectrum. The computational time is reduced even further if stellar parameters are known. However the accuracy of the stellar parameters is not important, because the spectra with vaguely similar parameters are compared each time. As soon as the number of observed spectra in a survey is large, the method can be used early in the data analysis, as it does not depend on the results of other pipelines.

Interstellar lines can serve as a valuable tracer of the interstellar extinction. Stars observed in large spectroscopic surveys often do not have photometric measurements made at the level where extinction can be calculated. The derived extinction can then be used to correct distance calculations. A large database of  DIB observations is very valuable among the DIB community, where individual stars are most often observed and large catalogues of DIBs consist of only a few thousands of sightlines. With more extensive observations the distribution of DIB carriers can be studied and 3D maps can be constructed. For RAVE observations, this aspect will be covered in our next paper.

We applied this new method to RAVE stars. Due to the low S/N of RAVE spectra, and due to the fact that only sky regions away from the Galactic plane were observed, the diffuse interstellar band at 8620 {\AA} could not be observed in single spectra, except for very few examples (spectra with \textbf{S/N$>$300} or stars in few regions with high reddening). Instead, different volumes were used, where spectra were combined to gain enough S/N and to be able to detect the DIB.

We measured the equivalent width for the DIB and compared it to reddening from \citet{schlegel98} and to reddening derived by a Bayesian method. A linear relation was fitted to set the relation between the reddening and the strength of the DIB from a large sample of observed sightlines. For \citet{schlegel98} reddening values we fitted an envelope and for Bayesian reddening we fitted an ordinary linear relation to the datapoints. Slopes of both relations are in agreement. When the offset for Bayesian reddening was accounted for, both relations also show that the linear relation goes trough the zero point. The slope is also in agreement with the linear relation for hot stars, measured without dividing by the stellar spectrum. 

One important assumption is made in our method namely, that there are stars observed with negligible reddening. This proved to be a safe assumption, because we do not see anything that would suggest that the equivalent widths are underestimated. The linear relation shows that the equivalent width of DIB 8620 drops to zero when the reddening is zero. This is not necessarily true for interstellar lines, but it is true at least for strong diffuse interstellar bands, especially in the regions that are not shielded from UV radiation \citep{kos13}. This is mostly the case for the observed regions.

Three methods described in this paper give matching relations between the equivalent width of DIB 8620 and the reddening, but they differ from the relations in the existing literature. Our slopes are $\sim$10~\% flatter than reported in the literature. We were not able to identify the reason for this discrepancy, though intrinsic differences in the sightlines toward individual hot stars with a relatively high reddening may be responsible.

\textbf{The method described in this paper will see further usage in the coming large spectroscopic surveys. A spectra of similar S/N ratio, but of higher resolution and in more bandpasses are already being taken for the Gaia-ESO survey \citep{gilmore12}. Many more DIBs will be detectable in Gaia-ESO spectra and will also cover different parts of the Galaxy, so a study of DIBs on a large scale and in different environments will be possible. A step further will be Hermes-GALAH survey \citep{galah12}, that will cover four different bandpasses, with a resolving power of up to 48,000 and with a much higher S/N. This will allow to observe DIBs in the spectra of individual stars. Also the number of observed stars is planed to be two times larger than in RAVE and 10 times larger than in Gaia-ESO. Most similar to RAVE spectra will be the spectra from the radial velocity spectrograph on board the Gaia mission \citep{gaia12}. They will cover the same spectral range with a similar resolution, but will have much lower S/N. However, the number of observed targets will be much larger and spectra could be combined in the same way as described in this paper.}

The offset of the Bayesian reddening was measured together with the scale height of the dust disk, giving approximately 120~pc. The scale height measured from the DIB compared to dust and pseudo 3D maps will be studied in detail in the next paper about DIB 8620 in RAVE.

\emph{Acknowledgements:}\\
Funding for RAVE has been provided by: the Anglo-Australian Observatory;
the Leibniz-Institut f\"ur Astrophysik Potsdam; the Australian National
University; the Australian Research Council; the French National Research
Agency; the German Research foundation; the Istituto Nazionale di
Astrofisica at Padova; The Johns Hopkins University; the National Science
Foundation of the USA (AST-0908326); the W.M. Keck foundation; the
Macquarie University; the Netherlands Research School for Astronomy; the
Natural Sciences and Engineering Research Council of Canada; the Slovenian
Research Agency; Center of Excellence Space.si; the Swiss National Science
Foundation; the Science \& Technology Facilities Council of the UK;
Opticon; Strasbourg Observatory; and the Universities of Groningen,
Heidelberg and Sydney. The RAVE web site is at
\href{http://www.rave-survey.org}{http://www.rave-survey.org}.\\

\bibliographystyle{apj}
\bibliography{dibbib}

\end{document}